\documentclass{llncs}

\usepackage{amsmath}
\usepackage{amsfonts}
\usepackage{amssymb}
\usepackage{graphicx}
\usepackage{multirow}
\usepackage[english]{babel}
\usepackage[square,numbers,sort&compress]{natbib}

\newcommand{\indep}{\perp\hspace{-0.21cm}\perp}
\newcommand{\given}{\mid}
\newcommand{\Prob}{\operatorname{P}}

\numberwithin{equation}{section}
\numberwithin{figure}{section}

\begin{document}

\title{Graphical Modelling in Genetics \\ and Systems Biology}
\titlerunning{Graphical Modelling in Systems Biology}
\author{Marco Scutari}
\authorrunning{Marco Scutari}
\tocauthor{Marco Scutari}
\institute{Genetics Institute, University College London, London, United Kingdom.}

\maketitle

Graphical modelling in its modern form was pioneered by Lauritzen and Wermuth
\cite{wermuth} and Pearl \cite{pearl} in the $1980$s, and has since found
applications in fields as diverse as bioinformatics \cite{gene}, customer
satisfaction surveys \cite{salini} and weather forecasts \cite{hailfinder}. 
Genetics and systems biology are unique among these fields in the dimension of
the data sets they study, which often contain several thousand variables and
only a few tens or hundreds of observations. This raises problems in both
computational complexity and the statistical significance of the resulting
networks, collectively known as the ``curse of dimensionality''. Furthermore,
the data themselves are difficult to model correctly due to the limited
understanding of the underlying phenomena. In the following, we will
illustrate how such challenges affect practical graphical modelling and some
possible solutions.

\section{Background and Notation}
\label{sec:notation}

Graphical models \cite{koller,pearl} are a class of statistical models composed
by a set $\mathbf{X} = \{X_1, X_2, \ldots, X_p\}$ of \emph{random variables}
describing the quantities of interest and a \emph{graph} $\mathcal{G} = 
(\mathbf{V}, E)$ in which each \emph{node} or \emph{vertex} $v \in \mathbf{V}$
is associated with one of the random variables in $\mathbf{X}$. The \emph{edges}
$e \in E$ are used to express direct dependence relationships among the variables
in $\mathbf{X}$. The set of these relationships is often referred to as the 
\emph{dependence structure} of the graph. Different classes of graphs express
these relationships with different semantics, which have in common the principle
that graphical separation of two vertices implies the conditional independence
of the corresponding random variables \cite{pearl}. Examples most commonly found
in literature are \emph{Markov networks} \cite{whittaker,edwards}, which use 
undirected graphs; \emph{chain graphs} \cite{cox}, which use partially directed
graphs; and \emph{Bayesian networks} \cite{neapolitan,korb}, which use directed
acyclic graphs.

In principle, there are many possible choices for the joint distribution of
$\mathbf{X}$, depending on the nature of the data and the aims of the analysis.
However, literature has focused mostly on two cases: the \emph{discrete
case} \cite{whittaker,heckerman}, in which both $\mathbf{X}$ and the $X_i$ are
multinomial random variables, and the \emph{continuous case} \cite{whittaker,heckerman3},
in which $\mathbf{X}$ is multivariate normal and the $X_i$ are univariate normal
random variables. In the former, the parameters of interest are the
\emph{conditional probabilities} associated with each variable, usually
represented as conditional probability tables; in the latter, the parameters
of interest are the \emph{partial correlation coefficients} between each
variable and its neighbours in $\mathcal{G}$.

The estimation of the structure of $\mathcal{G}$ is called \emph{structure
learning} \cite{koller,edwards}, and consists in finding the graph that encodes
the conditional independencies present in the data. Ideally it should coincide
with the dependence structure of $\mathbf{X}$, or it should at least identify a
distribution as close as possible to the correct one in the probability space.
Several algorithms have been presented in literature for this problem. Despite
differences in theoretical backgrounds and terminology, they can all be traced
to three approaches: \emph{constraint-based} (which are based on conditional
independence tests), \emph{score-based} (which are based on goodness-of-fit 
scores) and \emph{hybrid} (which combine the previous two approaches). For some
examples, see Castelo and Roverato \cite{roverato}, Friedman et al. \cite{sc},
Larra{\~n}aga et al. \cite{larranaga} and Tsamardinos et al. \cite{mmhc}. All
these structure learning algorithms operate under a set of common assumptions:
\begin{itemize}
  \item there must be a one-to-one correspondence between the nodes in the graph
    and the random variables in $\mathbf{X}$; this means in particular that
    there must not be multiple nodes which are deterministic functions of a
    single variable;
  \item observations must be independent. If some form of temporal or spatial
    dependence is present, it must be specifically accounted for in the definition
    of the network, as in \emph{dynamic Bayesian networks} \cite{koller};
  \item every combination of the possible values of the variables in $\mathbf{X}$
    must represent a valid, observable (even if really unlikely) event. 
\end{itemize}
On the other hand, the structure of the network can also be specified from prior
knowledge of the phenomenon underlying the data; in this case the graphical
model implements an \emph{expert system} \citep{cowell,castillo}. This is rarely
done in practice, especially in genetics and systems biology, because available
information are typically scarce or unreliable. It is far more common to use such 
information to inform the choices made by a structure learning algorithm, thus
making the best use of the data \cite{MS2008}.

The structure of a graphical model has two important properties. The first is
that it defines the decomposition the probability distribution of $\mathbf{X}$,
called the \emph{global distribution}, into a set of \emph{local distributions}.
For practical reasons, each local distribution should involve only a small number
of variables when applying graphical modelling to high dimensional problems. For 
Bayesian networks it is related to the chain rule of probability \citep{korb};
it takes the form
\begin{equation}
\label{eqn:parents}
  \Prob(\mathbf{X}) = \prod_{i=1}^p \Prob(X_i \given \Pi_{X_i})
\end{equation}
so that each local distribution is associated with a single node $X_i$ and
depends only on the joint distribution of its parents $\Pi_{X_i}$. This
decomposition holds for any Bayesian network, regardless of its graph structure.
In Markov networks local distributions are associated with the \textit{cliques}
$\mathbf{C}_1$, $\mathbf{C}_2$, $\ldots$, $\mathbf{C}_k$, the maximal subsets
of nodes in which each element is adjacent to all the others:
\begin{equation}
\label{eqn:cliques}
  \Prob(\mathbf{X}) = \prod_{i=1}^k \psi_i(\mathbf{C}_i).
\end{equation}
The functions $\psi_1, \psi_2, \ldots, \psi_k$ are called \textit{Gibbs'
potentials} \citep{pearl}, \textit{factor potentials} \citep{castillo} or simply
\textit{potentials}, and are non-negative functions representing the relative
mass of probability of each clique. They are proper probability or density 
functions only when the graph is \textit{decomposable} or \textit{triangulated},
that is, when it contains no induced cycles other than triangles. In this case
the global distribution factorises again according to the chain rule and can
be written as
\begin{equation}
  \Prob(\mathbf{X}) = \frac{\prod_{i=1}^k \Prob(\mathbf{C}_i)}{\prod_{i=1}^k \Prob(\mathbf{S}_i)}
\end{equation}
where $\mathbf{S}_i$ are the nodes of $\mathbf{C}_i$ which are also part
of any other clique up to $\mathbf{C}_{i-1}$ \citep{pearl}.

The second important property is that the \emph{Markov blanket} of each node
can be easily identified from the structure of the graph. For instance, in
Bayesian networks the Markov blanket of a node $X_i$ is the set consisting of
the parents of $X_i$, the children of $X_i$ and all the other nodes sharing
a child with $X_i$ \cite{pearl}. Since the Markov blanket is defined as the
set of nodes that makes the target node (\emph{i.e.} $X_i$) independent from
all the other nodes in $\mathbf{X}$, it provides a theoretically-sound
solution to the \emph{feature selection} problem \cite{feature}.

\section{Data and Models in Statistical Genetics and Systems Biology}
\label{sec:datatypes}

In genetics and systems biology, graphical models are employed to describe and
identify interdependencies among genes and gene products, with the eventual
aim to better understand the molecular mechanisms linking them. Data made
commonly available for this task by current technologies fall into three
groups: 
\begin{enumerate}
  \item gene expression data \cite{gene,spirtes2}, which measure the intensity of
    the activity of a particular gene through the presence of \emph{messenger
    RNA} (mRNA, for protein-coding genes) or other kinds of \emph{non-coding RNA}
    (ncRNA, for non-coding genes);
  \item protein signalling data \cite{sachs}, which measure the proteins produced
    as a result of each gene's activity;  
  \item sequence data \cite{gianola}, which provide the nucleotide sequence of
    each gene. For both biological and computational reasons, such data contain
    mostly \emph{single-nucleotide polymorphisms} (SNPs) -- genes which vary in
    only one nucleotide between individuals -- having only two possible alleles,
    called \emph{biallelic SNPs}.
\end{enumerate}

In the case of gene expression and protein signalling data (Sections 
\ref{sec:expression} and \ref{sec:protein}), we are interested in grouping them
into tempporal sequences determining some molecular process (the \emph{functional
pathways}). Bayesian networks are naturally suited to this task. If we assign
each gene to one node in the network, edges represent the interplay between 
different genes. They can describe either direct interactions or indirect 
influences that are mediated by unobserved genes. This is a crucial property
because it is impossible in practice to completely observe a complex molecular
process: either we do not know all the genes involved or we may be unable to
obtain reliable measurements of all their expression levels. Furthermore, under
appropriate conditions \cite{causality,koller} edge directions may be indicative
of the causal relationships in the underlying pathways. In that case, the 
Bayesian network reflects the ordering of connections between pathway components
and the actual flow of the molecular process.

Similar considerations can be made when protein signalling data are used just
to identify protein-protein interactions, limiting ourselves to the study of
the cell's physiology.

On the other hand, in sequence data analysis (Section \ref{sec:snp}) we are 
interested in modelling the behaviour of one or more \emph{phenotypic traits}
(\emph{e.g.} the presence of a disease in humans, yield in plants, milk 
production in cows, etc.) by capturing direct and indirect causal genetic
effects. Unless some prior knowledge on the \emph{genetic architecture} of a
trait is available, a large set of genes spread over the whole genome is 
required for such effects to be detectable. If the focus is on identifying
the genes that are strongly associated with a trait, the analysis is called a 
\emph{genome-wide association study} (GWAS).

Applications of Bayesian networks to sequence data are more problematic than
in the previous cases; some care must be taken in their interpretation as
causal models. Edges linking genes to a trait can be considered direct
associations. As was the case for gene expression data, under appropriate
conditions such associations may actually be indicative of real causal effects.
On the other hand, edges linking genes to other genes arise from the genetic
structure of the individuals in the sample. It is expected, for example, that
genes that are located near each other on a chromosome are more likely to be
inherited together during meiosis, and are therefore said to be \emph{genetically
linked} \cite{falconer}. Furthermore, even genes that are far apart in the
genome can be in \emph{linkage disequilibrium} (LD) if some of their
configurations occur more often or less often than it would be expected from
their marginal frequencies. Both these phenomena induce associations between
the genes, but not cause-effects relationships. From a strictly causal point
of view, a chain graph in which genes are linked by undirected edges and the
only directed edges are the ones incident on the traits provides a better
visual representation of the network structure.

\subsection{Gene Expression Data}
\label{sec:expression}

Gene expression data are typically composed of a set of \emph{intensities}
measuring the abundance of several RNA patterns, each meant to probe a particular
gene. These intensities are measured either radioactively or fluorescently,
using labels that mark the desired RNA patterns \cite{duggan,lennon,lipshutz}. 

The measured abundances present several limitations. First of all, microarrays
measure abundances only in terms of relative probe intensities, not on an
absolute scale. As a result, comparing different studies or including them in a
meta-analysis is difficult in practice without the use of rank-based methods
\cite{breitling}. Furthermore, even within a single study abundance measurements
are systematically biased by \emph{batch effects} introduced by the instruments
and the chemical reactions used in collecting the data \cite{schuchhardt}.

\begin{figure}[t]
\begin{center}
  \includegraphics[width=0.8\textwidth]{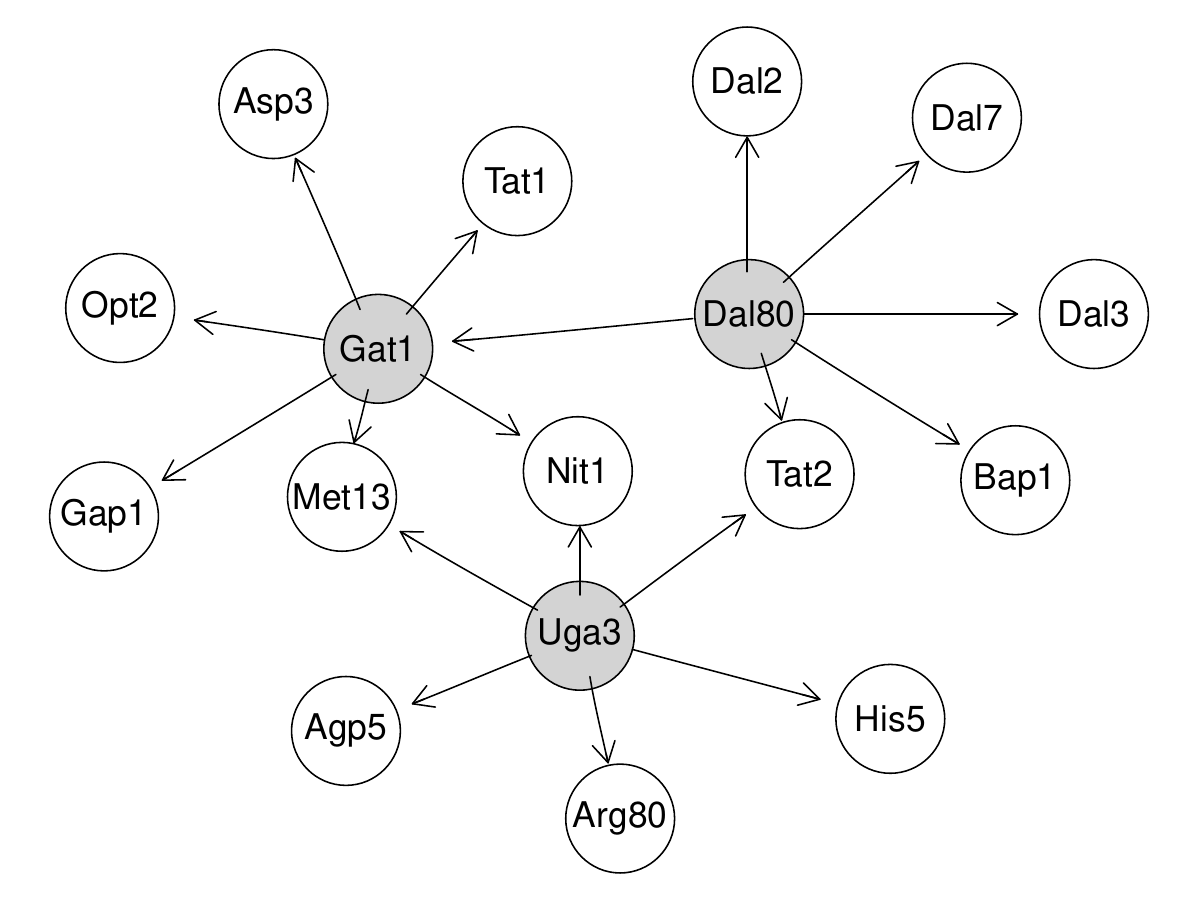}
  \caption{A Bayesian network learned from gene expression data and used as an
   example in Friedman \cite{Fri04}. Grey nodes correspond to the \emph{regulators}
   of the  network, the genes controlling the expression of the other (\emph{target})
   genes involved in a molecular process.}
  \label{fig:friedman}
\end{center}
\end{figure}

By their nature, gene expression data are modelled as continuous random variables
and are investigated using Pearson's correlation, either assuming a Gaussian
distribution or applying results from robust statistics \cite{huber,thomas}. 
The simplest graphical models used for gene expression data are \emph{relevance
networks} \cite{BTS+00}, also known in statistics as \emph{correlation graphs}.
Relevance networks are constructed by estimating the correlation matrix of the
genes and thresholding its elements, so that weak correlations are set to zero.
Finally, a graph is drawn in order to depict the remaining strong correlations.

\emph{Covariance selection models} \cite{Dem72}, also known as \emph{concentration
graphs} or \emph{graphical Gaussian models} \cite{whittaker}, consider conditional
rather than marginal dependencies; the presence of an edge is determined by the
value of the corresponding \emph{partial correlation}. In the context of systems
biology, the resulting graphs are often called \emph{gene association networks},
and are not trivial to estimate from high-dimensional genomic data. Several 
solutions have been proposed in literature, based either on James-Stein 
regularisation \cite{SS05a,SS05c} or on different penalised maximum likelihood
approaches \cite{LG06,BGA2008,FHT2008}.

Both gene relevance and gene association networks are undirected graphs. The
application of Bayesian networks to learn large-scale directed graphs from
microarray data was pioneered by Friedman \emph{et al.} \cite{FLNP00}, and has
also been reviewed more recently in Friedman \cite{Fri04} (see Figure 
\ref{fig:friedman}). The high dimensionality of the model, combined with low 
sample sizes, means that inference procedures are usually unable to identify a
single best Bayesian network, settling instead on a set of equally well behaved
models. For this reason, it is important to incorporate prior biological 
knowledge into the network through the use of informative priors \cite{MS2008}
and to produce confidence scores in its graphical features \cite{imoto,friedman}.

\subsection{Protein Signalling Data}
\label{sec:protein}

\begin{figure}[t]
\begin{center}
  \includegraphics[width=0.8\textwidth]{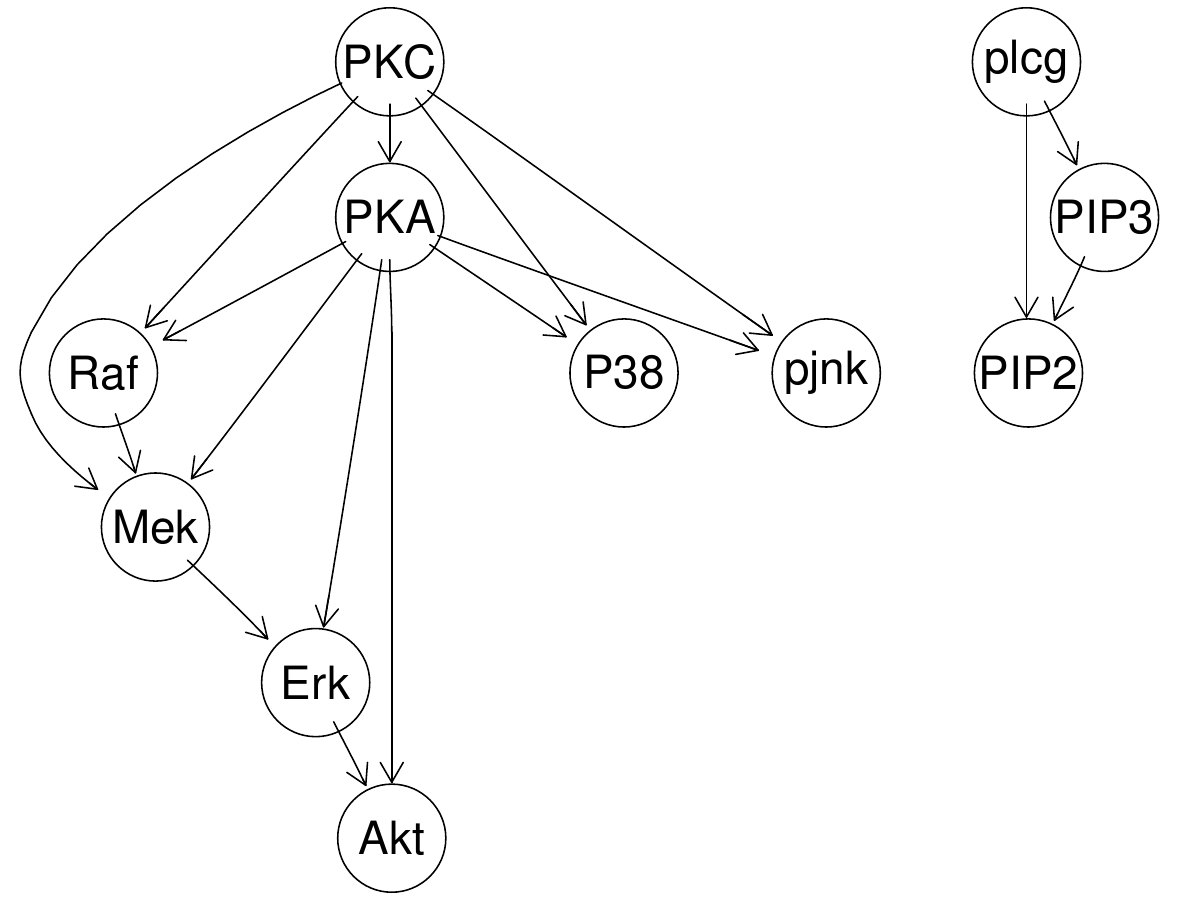}
  \caption{The Bayesian network learned from the protein-signalling data in
    Sachs \emph{et al.} \cite{sachs} using model averaging and data from
    several experiments performed under different stimulatory and inhibitory
    cues.}
  \label{fig:sachs}
\end{center}
\end{figure}

Protein signalling data are similar to gene expression data in many respects,
and in fact are often used to indirectly investigate the expression of a set
of genes. In general, the relationships between proteins are indicative of
their physical location within the cell and of the development over time of
the molecular processes they are involved in.
 
From a modelling perspective, all the approaches covered in Section 
\ref{sec:expression} can be applied to protein signalling data with little or
no change. However, it is important to note that protein signalling data 
sometimes have sample sizes that are much larger than either gene expression
or sequence data; an example is the study from Sachs \emph{et al.} \cite{sachs}
on how to derive a causal Bayesian network from multi-parameter single-cell
data (Figure \ref{fig:sachs}).

\subsection{Sequence Data}
\label{sec:snp}

Sequence data are fundamentally different from both gene expression and protein
signalling data, for several reasons. First, sequence data provide direct access
to the genome's information, without relying on indirect measurements. As a
result, they provide a closer view of the genetic layout of an organism than other
approaches. Second, sequence information is intrinsic to each individual, and
does not vary over time; therefore, the inability of static Bayesian networks
to model feedback loops is not a limitation in this case. 

Furthermore, sequence data is naturally defined on a discrete rather than
continuous domain. Each gene has a finite number of possible states, determined
by the number of combinations of nucleotides differing between the individuals
in the sample. In the case of biallelic SNPs, each SNP $X_i$ differs at a single
base-pair location and has only three possible variants. They are determined by
the (unordered) combinations of the two nucleotides observed at that location,
called the \emph{alleles}, and are often denoted as ``AA'', ``Aa'', ``aa''. The
``A'' and ``a'' labels can be assigned to the nucleotides in several ways; for
instance, ``A'' can be chosen as either the most common in the sample (which
makes models easier to interpret) or by following the alphabetical order of the 
nucleotides (which makes the labelling independent from the sample). ``AA''
and ``aa'' individuals are said to be \emph{homozygotes}, because both nucleotides
in the pair have the same allele; ``Aa'' individuals are said to be
\emph{heterozygotes}.

From a graphical modelling perspective, modelling each SNP as a discrete
variable is the most convenient option; multinomial models have received much
more attention in literature than Gaussian or mixed ones. On the other hand,
the standard approach in genetics is to recode the alleles as
numeric variables, \emph{e.g.}
\begin{align}
\label{eq:alleles}
  &X_i = \left\{
    \begin{aligned}
      1&  \text{ if the SNP is ``AA''} \\
      0&  \text{ if the SNP is ``Aa''} \\ 
      -1& \text{ if the SNP is ``aa''} 
    \end{aligned}
  \right.&
  &\text{or}&
  &X_i = \left\{
    \begin{aligned}
      2&  \text{ if the SNP is ``AA''} \\
      1&  \text{ if the SNP is ``Aa''} \\ 
      0&  \text{ if the SNP is ``aa''} 
    \end{aligned}
  \right.\,.
\end{align}
In both cases, the recoded variables are typically modelled using an additive 
Bayesian linear regression model of the form
\begin{align}
\label{eq:additive}
  &\mathbf{y} = \mu + \sum_{i = 1}^n X_i g_i + \boldsymbol{\varepsilon},&
  &g_i \sim \pi_{g_i},\,\boldsymbol{\varepsilon} \sim N(\mathbf{0}, \mathrm{\Sigma})
\end{align}
where $g_i$ denotes the effect of gene $X_i$, $\mathbf{y}$ is the trait under
study and $\mu$ is the population mean. The matrix $\mathrm{\Sigma}$ models the
relatedness of the subjects, which is called \emph{kinship} in genetics, and
populations structure \cite{astle}. In human genetics, it is often assumed to
be the identity matrix, which implies the assumptions that individuals are
unrelated. Several implementations of Equation \ref{eq:additive} based on
linear mixed models and penalised regressions have been proposed, mostly within
the framework of Bayesian statistics. Some examples are the Genomic BLUP (GBLUP),
BayesA and BayesB from Meuwissen et al. \cite{meuwissen}, the Bayesian LASSO
from Park and Casella \cite{blasso} and the BayesC$\pi$ from Habier et al.
\cite{bayescpi}.

\begin{figure}[t]
\begin{center}
  \includegraphics[width=0.8\textwidth]{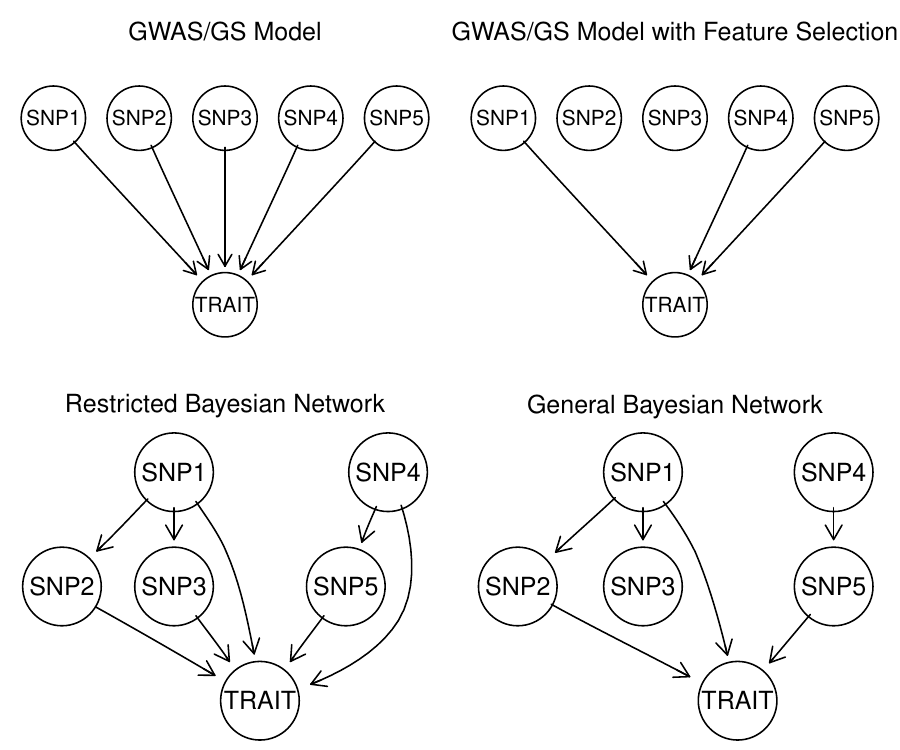}
  \caption{Different approaches to GWAS. On the top, classic additive Bayesian
    linear regression models with and without feature selection. On the bottom,
    more complex models based on Bayesian networks.}
  \label{fig:bn-gwas}
\end{center}
\end{figure}

Graphical models, and Bayesian networks in particular, provide a systematic way
to categorise and extend such models. Consider the four different models shown
in Figure \ref{fig:bn-gwas}. The classic additive model from Equation 
\ref{eq:additive} is shown in the top-left panel; SNPs are independent from each
other and all contribute in explaining the behaviour of the phenotypic trait.
This is the case for BayesA and GBLUP. In the top-right panel, some SNPs are
identified as non-significant and excluded from the additive model. Models of
this kind include BayesB, BayesC$\pi$ and the Bayesian LASSO, which perform
feature selection in the context of model estimation.

A natural way to extend these models is to include interactions between the
SNPs, as shown in the two bottom panels. A recent study by Morota \emph{et al.}
\cite{gianola} has shown that assuming additive effects can only be justified
on the grounds of computational efficiency, because interactions between the SNPs
are so complex that even pairwise dependence measures are not able to capture
them completely. On the other hand, Bayesian networks provide a more accurate
picture of these dependencies and are more effective at capturing and displaying
them. If the trait is discrete, Bayesian network classifiers \cite{tan} such
as the Tree-Augmented Na\"ive Bayes (TAN) can also be used to implement GWAS
models.

\section{Challenges in Bayesian Network Modelling}

Gene expression, protein signalling and sequence data are difficult to analyse
in a rigorous and effective way regardless of the model used, as they present
significant computational and statistical challenges. We review some of them in
the following, concentrating on those that affect the earliest stages of model
specification. Obviously, the quality of the models estimated from the data
rests crucially on their structure and estimation; and the accuracy of 
subsequent inference may vary substantially depending on how model specification
relates to the phenomena under investigation. 

The combination of small sample sizes and large numbers of variables ($n \ll p$),
often called the ``curse of dimensionality'', is perhaps the most evident problem
in model specification and algorithm implementation. This is especially true for 
Bayesian networks, because both learning and inference are NP-hard 
\cite{npcomplete,nphard}. This may rise some concerns about the amount of 
information present in the data and in the computational complexity of model
estimation (Section \ref{sec:npsmall}). The former can be tackled by effective
distributional assumptions (Section \ref{sec:ordinal}), and the latter by the
use of feature selection to reduce the dimensionality of the problem (Section
\ref{sec:feature}).

\subsection{Limits of the ``$n \ll p$'' Data Sets}
\label{sec:npsmall}

The disparity between the available sample sizes and the number of genes or
proteins under investigation is probably the most important limiting factor in
genetics and systems biology. In a few cases, the underlying phenomenon is
known to the extent that only the relevant variables are included in the model
(Sachs \emph{et al.} \cite{sachs} is one such study). However, in general 
molecular processes are so complex that statistical modelling is used more as
a tool for exploratory analysis than to provide mechanistic explanations. In
the former case, we have that $n \gg p$, and we can use results from large-sample
theory \cite{lehmann2} and computationally-intensive techniques \cite{comp,comp2}
in selecting and estimating our models. In the latter, the limits of the model
depend heavily on what knowledge is available on the phenomenon and on our
ability to incorporate it in the prior.

Consider, following Bayes' theorem, the posterior distribution of the parameters 
in the model (say $\boldsymbol{\theta}$) given the data
\begin{equation}
  p(\boldsymbol{\theta} \given \mathbf{X}) \propto 
    p(\mathbf{X} \given \boldsymbol{\theta}) \cdot p(\boldsymbol{\theta}) =
    L(\boldsymbol{\theta}; \mathbf{X}) \cdot p(\boldsymbol{\theta})
\end{equation}
or, equivalently,
\begin{equation}
  \log p(\boldsymbol{\theta} \given \mathbf{X}) = 
     c + \log L(\boldsymbol{\theta}; \mathbf{X}) + \log p(\boldsymbol{\theta}).
\end{equation}
The log-likelihood, $\log L(\boldsymbol{\theta}; \mathbf{X})$, is a function of
the data and therefore scales with the sample size, while the prior density does
not. For small sample sizes, there may not be enough data available to disprove
the assumptions encoded in the prior. As a result, conclusions arising from model
estimation and inference reflect our beliefs on the phenomenon (as encoded in
the prior) more than the reality of the observed molecular processes. In this
context, even the use of non-informative priors may result in posteriors with
undesirable properties \cite{bernardo}.

In that regard, Bayesian networks present considerable advantages. First, they
are very flexible in specifying variable selection rates and interactions. In
other words, the prior makes fewer assumptions on the probabilistic structure
of the data and is therefore less likely to completely dominate the likelihood.
Second, the effects of the values assigned to the parameters of a non-informative
prior are well understood for both small and large samples \cite{steck1,steck2},
and corrected posterior density functions are available in closed form. 

Another important consideration is the ease of estimating the model. Models used
in genetics and systems biology often require expensive Markov Chain Monte Carlo
simulations; two such examples are BayesA and BayesB. On the other hand, many
closed form results are available for both discrete and Gaussian Bayesian networks.
For networks up to $100$ variables, exact structure learning algorithms are
available \cite{koivisto} and exact inference algorithms such as Variable
Elimination and Clique Trees \cite{koller} are feasible to use. For larger
networks, efficient structure learning heuristics such as the  Semi-Interleaved
Hiton-PC from Aliferis \emph{et al.} \cite{hiton1,hiton2} and approximate
inference algorithms such as the Adaptive Importance Sampling for Bayesian
Networks (AIS-BN) from Cheng and Druzdel \cite{aisbn} are feasible up to several
thousand variables.

\subsection{Discrete or Continuous Variables?}
\label{sec:ordinal}

All the data types covered in Section \ref{sec:datatypes} are often modelled
using Gaussian Bayesian networks, which represent the natural evolution of the
linear regression models used in literature. In the case of gene expression and
protein signalling data, sometimes \cite{sachs,hartemink} the data are 
discretised into intervals and a discrete Bayesian network is used instead. As
for gene expression data, both Gaussian and discrete Bayesian networks can be
used depending on whether we use the numeric coding in Equation \ref{eq:alleles}
or not.

Clearly, both distributional assumptions present important limitations. Gaussian
Bayesian networks assume that the global distribution is multivariate normal.
This is unreasonable in the case of sequence data, which can only assume a finite,
discrete set of values. Gene expression and protein signalling data, while
continuous, are in general significantly skewed unless preprocessed with a Box-Cox
transformation \cite{box-cox}. Furthermore, Gaussian Bayesian networks are only
able to capture linear dependencies, and have a low power in detecting non-linear
ones. On the other hand, using discrete Bayesian networks and assuming a multinomial
distribution disregards useful information present in the data and may result in 
models with a very large number of parameters. If the ordering
of the intervals (in discretised gene expression and protein signalling data)
or of the alleles (in sequence data) is ignored, both learning and subsequent
inference are not aware that dependencies are likely to take the form of
stochastic trends. This is true, in particular, for sequence data, as the effect
of the heterozygous allele is necessarily comprised between the effect of the
two heterozygous alleles. 

An approach that has the potential to outperform both discrete and Gaussian
assumptions has been recently proposed by Musella \cite{musella} with Bayesian
networks learned from ordinal data. Structure learning is performed with a 
constraint-based approach (in particular, the PC algorithm from Sprites
\emph{et al.} \cite{spirtes}) using the Jonckheere-Terpstra test for trend
among ordered alternatives \cite{jonckheere,terpstra}. Consider a conditional
independence test for $X_1 \indep X_3 \given X_2$, where $X_1$, $X_2$ and $X_3$
have $T$, $L$ and $C$ levels respectively. The test statistic is defined as
\begin{equation}
  JT = \sum_{k = 1}^{L} \sum_{i = 2}^{T} \sum_{j = 1}^{i - 1} 
         \left[ \sum_{s = 1}^{C} w_{ijsk} n_{isk} - \frac{n_{i+k}(n_{i+k} + 1)}{2}\right]
\end{equation}
where the $w_{ijsk}$ are Wilcoxon scores, defined as
\begin{equation}
  w_{ijsk} = \sum_{t = 1}^{s - 1} \left[ n_{itk} + n_{jtk} + \frac{n_{isk} + n_{jsk} + 1}{2}\right],
\end{equation}
and has an asymptotic normal distribution with mean and variance defined in
Lehmann \cite{lehmann} and Pirie \cite{pirie}. The null hypothesis is that of
homogeneity; if we denote with $F_{i,k}(x_3)$ the distribution function of
$X_3 \given X_1 = i, X_2 = k$, 
\begin{align*}
  &H_0: F_{1,k}(x_3) = F_{2,k}(x_3) = \ldots = F_{T,k}(x_3)& &\text{for $\forall x_3$ and $\forall k$}.
\end{align*}
The alternative hypothesis $H_1 = H_{1,1} \cup H_{1,2}$ is that of stochastic
ordering, either increasing
\begin{align*}
  &H_{1,1}: F_{i,k}(x_3) \geqslant F_{j,k}(x_3)& &\text{with $i < j$ for $\forall x_3$ and $\forall k$}
\end{align*}
or decreasing
\begin{align*}
  &H_{1,2}: F_{i,k}(x_3) \leqslant F_{j,k}(x_3)& &\text{with $i < j$ for $\forall x_3$ and $\forall k$}.
\end{align*}

\begin{figure}[t]
\begin{center}
  \includegraphics[width=0.9\textwidth]{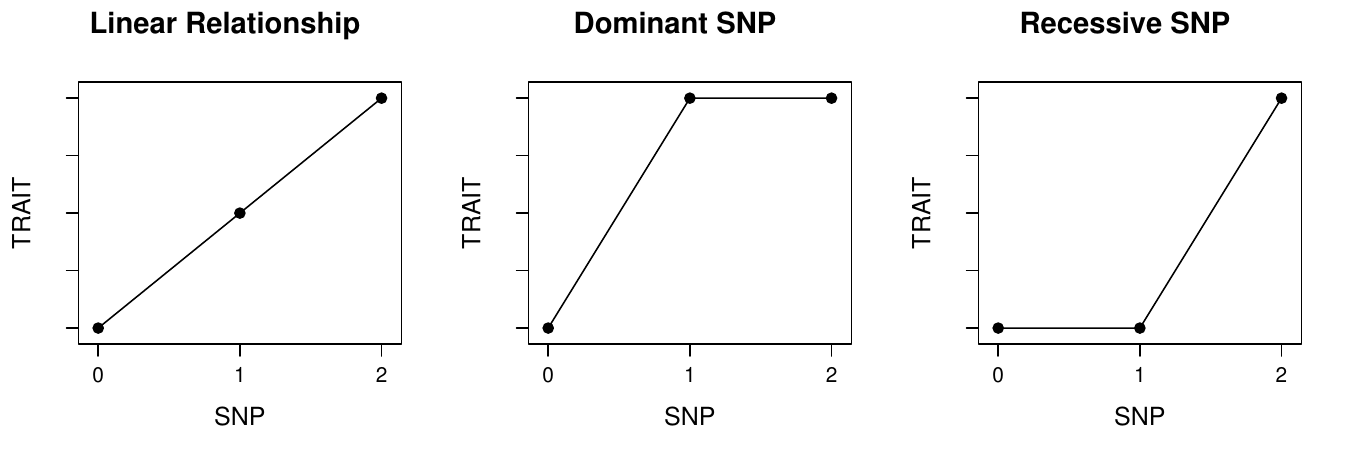}
  \caption{Three patterns of SNP effects on a phenotypic trait: linear association
    (left), a dominant SNP (centre), a recessive SNP (right).}
  \label{fig:dominance}
\end{center}
\end{figure}

The advantages of the Jonckheere-Terpstra test compared to linear association
can be illustrated, for example, by considering the different patterns of 
dominance of a single SNP shown in Figure \ref{fig:dominance}. Due to the way
SNPs are recoded as numeric variables, assuming that dependence relationships
are linear (left panel) forces the effect of heterozygotes to be the mean of
the effects of the the respective homozygotes. This is not always the case, as
SNPs can be \emph{dominant} (centre) or \emph{recessive} (right) for a trait,
either singly or in groups \cite{falconer}. Tests for linear association have
very low power against such nonlinear alternative hypotheses. On the other
hand, the alternative hypothesis of the Jonckheere-Terpstra test characterises
correctly both dominant and recessive SNPs. Furthermore, the Jonckheere-Terpstra
test exhibits more power than the independence tests used in discrete Bayesian
networks because of the more specific alternative hypothesis (\emph{e.g.}
stochastic ordering is just one particular case of stochastic dependence).

\subsection{Feature Selection as a Data Pre-Processing Step}
\label{sec:feature}

It is not possible, nor expected, for all genes in modern, genome-wide data sets
to be relevant for the trait or the molecular process under study. In part, this
is because of the curse of dimensionality, but it is also because different genes 
may provide essentially the same information due to linkage disequilibrium.
Furthermore, the effects of some genes on a trait may be mediated by other genes,
thus making them redundant. For this reason, in practice statistical models in
systems biology and genetics require a feature selection to be performed, either
during the learning process or as a separate data pre-processing step. 

In the context of GWAS models, we aim to find the subset of genes $\mathbf{S}
\subset \mathbf{X}$ such that
\begin{equation}
  \Prob(\mathbf{y} \given \mathbf{X}) = 
    \Prob(\mathbf{y} \given \mathbf{S}, \mathbf{X} \setminus \mathbf{S}) 
    \approx \Prob(\mathbf{y} \given \mathbf{S}), 
\end{equation}
that is, the subset of genes ($\mathbf{S}$) that makes all other markers
\mbox{($\mathbf{X} \setminus \mathbf{S}$)} redundant as far as the trait
$\mathbf{y}$ we are studying is concerned. Markov blankets identify such a
subset in the framework of graphical models; several algorithms have been
proposed in literature for their learning \cite{iamb,hiton1}. After the set
$\mathbf{S}$ has been identified, we can either fit one of the Bayesian linear
regression models from Section \ref{sec:snp} or learn a Bayesian network from
$\mathbf{y}$ and $\mathbf{S}$. In both cases, the smaller number of variables
included in the model reduces the effects of the curse of dimensionality 
\cite{sagmb12}. On the other hand, the conditional independence tests used by
Markov blanket learning algorithms do not take kinship into account. Therefore,
interpreting edges from $\mathbf{S}$ to $\mathbf{y}$ as direct causal influences
may lead to spurious results, even when the model shows good predictive power
\cite{hiton1}.

As far as gene expression and protein signalling data are concerned, the problem
of feature selection is more complicated. In many cases, we are interested in
a complex molecular process, as opposed to a single trait. If we don't know
a priori at least some of the genes involved in the molecular process, performing
feature selection as a data pre-processing step is impossible; we have to identify
the pathways we are interested in from the structure of the Bayesian network
learned from $\mathbf{X}$. At most we can enforce sparsity in the network by
using shrinkage tests \cite{pesarin10} or non-uniform structural priors
\cite{friedman2}.

Even if we know which genes are involved, using Markov blankets for feature
selection presents significant drawbacks. The Markov blanket of each gene must
be learned separately because almost all algorithms in literature accept only
one target node. If no information is shared between different runs of the 
learning algorithm, this task is embarrassingly parallel but still computationally
intensive. If, on the other hand, we use backtracking and other optimisations
to share information between different runs, significant speed-ups are possible
at the cost of an increased error rate (\emph{i.e.} false positives and false
negatives among the nodes included in each Markov blanket). In both cases,
merging the Markov blankets of each gene into a single set requires the use of
\emph{symmetry corrections} \cite{mmhc,hiton1} that violate the proofs of
correctness of the learning algorithms.

\section{Conclusions}

Data sets in genetics and systems biology often contain several thousand
variables and only a few tens or hundreds of observations. This raises problems
in both computational complexity and the statistical significance of the
resulting networks, which are collectively known as the ``curse of dimensionality''.
Furthermore, the data themselves are difficult to model correctly due to the
limited understanding of the underlying molecular mechanisms. Bayesian networks
provide a very flexible framework to model such data, extending, complementing
or replacing classic models present in literature. Their flexibility in
incorporating prior knowledge, different parametric assumptions and different
dependence structures makes them a suitable choice for the analysis of gene
expression, protein signalling and sequence data.

\end{document}